\documentclass[twocolumn,prl,floatfix,preprintnumbers,a4paper,nofootinbib,superscriptaddress,groupedaddress]{revtex4-1}
\usepackage[english]{babel}
\usepackage[autostyle, english = american]{csquotes}
\MakeOuterQuote{"}
\usepackage[utf8]{inputenc}
\usepackage{amsmath,amssymb}
\usepackage{amsfonts}
\usepackage{bm}
\usepackage{graphics}
\usepackage{comment}
\usepackage{float}
\usepackage{amsmath}
\usepackage{amssymb}
\usepackage[normalem]{ulem} %for \sout
\usepackage[mathscr]{euscript}
\usepackage{acronym}
\usepackage{float}
\usepackage[caption=false]{subfig}
\usepackage{lipsum}
\usepackage{mathrsfs}
\usepackage{mathtools}
\usepackage{multirow}
\usepackage{graphicx}
\usepackage{mathtools}
\usepackage{multirow}
\usepackage{graphicx}
\usepackage[usenames,dvipsnames,table,xcdraw]{xcolor}
\usepackage[colorlinks, urlcolor=blue,citecolor=blue,pdfborder={0 0 0}]{hyperref} 
\usepackage[nameinlink,capitalise]{cleveref}
\usepackage{xspace}
\usepackage{xfp}
\usepackage{array}

\newcolumntype{C}[1]{>{\centering\arraybackslash}p{#1}}

\newcommand{\reply}[1]{{#1}}
\newenvironment{replies}{}{}

\definecolor{veryperi}{RGB}{102, 103, 171}
\definecolor{mygreen}{rgb}{0., 0.5, 0.}

\graphicspath{{fig}}

\begin{document}

\title{Neutrino Flavor Transformation in Neutron Star Mergers}

\author{Yi Qiu$^{1,2}$, David Radice$^{1,2,3}$, Sherwood Richers$^4$, and Maitraya Bhattacharyya$^{1,2}$}

\affiliation{$^1$ Institute for Gravitation and the Cosmos, The Pennsylvania State University, University Park PA 16802, USA}
\affiliation{$^2$Department of Physics, The Pennsylvania State University, University Park PA 16802, USA}
\affiliation{$^3$Department of Astronomy \& Astrophysics, The Pennsylvania State University, University Park PA 16802, USA}
\affiliation{$^4$Department of Physics \& Astronomy, University of Tennessee Knoxville, Knoxville TN 37996, USA}

\begin{abstract}
We present the first numerical relativity simulations including neutrino flavor transformations that could result from flavor instabilities, quantum many-body effects, or potential beyond standard model physics in neutron star mergers. We find that neutrino flavor transformations impact the composition and structure of the remnant, potentially leaving an imprint on the post-merger gravitational-wave signal. They also have a significant impact on the composition and nucleosynthesis yields of the ejecta.
\end{abstract}
%%%%%%%%%%%%%%%%%%%%%%%%%%%%%%%%%%%%%%%
\maketitle
%%%%%%%%%%%%%%%%%%%%%%%%%%%%%%%%%%%%%%%%
%%%%%%%%%%%%%%%%%%%%%%%%%%%%%%%%%%%%%%%%%%
\label{sec:intro}
\textit{Introduction.---}Multi-messenger astronomy is entering an exciting new era, with cutting-edge detectors set to observe gravitational waves (GWs) and their electromagnetic counterparts from compact object mergers at unprecedented rates~\cite{LIGOScientific:2016aoc, LIGOScientific:2017vwq, LIGOScientific:2017zic, KAGRA:2013rdx, Eichler:1989ve, Radice:2020ddv, Perego:2017wtu, Ruffert:1998qg, Just:2015dba, Cusinato:2021zin,Martin:2017dhc,Diamond:2023cto,Vigna-Gomez:2023euq,Most:2023sft,Hajela:2021faz,Rosswog:2024vfe}. This advancement promises to shed light on some of the most profound questions in astrophysics, particularly regarding the nature of dense matter in neutron stars~\cite{Baiotti:2016qnr, Baym:2017whm, Burgio:2021vgk,Janka:2025tvf}. While observations of kilonovae have confirmed that binary neutron star (BNS) mergers are a site of rapid neutron-capture (r-process) enrichment~\cite{Qian:1996xt,Hoffman:1996aj,Wu:2016pnw, Balantekin:2023ayx}, the exact yields and detailed composition of the ejecta remain uncertain~\cite{Kasen:2017sxr, Wanajo:2014wha,Zhu:2020eyk,Barnes:2020nfi,Foucart:2024kci}. These uncertainties stem from the complex interplay of nuclear physics, neutrino interactions~\cite{Espino:2023dei,Espino:2023mda}, and the hydrodynamical evolution of the merger remnant~\cite{Espino:2022mtb,Zenati:2024pgn}. In particular, neutrino flavor oscillations, which alter the electron fraction ($Y_e$) of the outflows, play a crucial role in shaping the conditions for r-process element synthesis~\cite{Zhu:2016mwa, Wu:2017drk,Li:2021vqj,Just:2022flt, Fernandez:2022yyv, Zenati:2023lwh}. 

Accurate modeling of these complex phenomena requires not only solving general relativistic magnetohydrodynamics (GRMHD) equations but also implementing a comprehensive scheme to evolve the neutrino radiation transport~\cite{Rosswog:1998hy,Foucart:2022bth,Collins:2022ocl,Volpe:2023met,Kawaguchi:2024naa}. In numerical simulations, modeling neutrino transport typically requires solving a 7-dimensional Boltzmann equation to capture their evolution in both coordinate and phase space. However, this approach remains computationally infeasible due to its exorbitant cost~\cite{Sumiyoshi:2020bdh,Bhattacharyya:2022bzf}. As a result, alternative methods such as the nondeterministic Monte Carlo approach~\cite{Foucart:2017mbt} and approximate schemes like truncated moment methods~\cite{Thorne:1981nvt, Shibata:2011kx, Cardall:2013kwa, Foucart:2016rxm, Sekiguchi:2015dma, Sekiguchi:2016bjd, Radice:2021jtw} are commonly employed~\cite{Foucart:2024npn,Cheong:2024buu}.

So far, binary neutron star merger simulations have neglected neutrino flavor conversions. Based on the neutrino distributions from simulations without flavor transformation, the flavor evolution under neutrino-neutrino interactions~\cite{Sigl:1993ctk,Sawyer:2005jk,Duan:2005cp, Izaguirre:2016gsx, Chakraborty:2016lct, Capozzi:2017gqd, Richers:2021xtf, Morinaga:2021vmc, Richers:2022zug,Wu:2021uvt,Nagakura:2021hyb,Capozzi:2022slf, Volpe:2023met, Abbar:2023zkm,Zaizen:2023wht} can occur on nanosecond timescales, which are orders of magnitude shorter than the timescales of hydrodynamic evolution, making them unfeasible to capture on the fly~\cite{Johns:2023jjt,Johns:2024dbe}. Important exceptions are the studies of post-merger accretion disks~\cite{Li:2021vqj, Just:2022flt, Fernandez:2022yyv}. However, they considered only the evolution of disks around compact objects and did not consider the merger phase itself, which is the most dynamical phase of the binary evolution and that sets the disk initial structure and composition~\cite{Camilletti:2024otr}. Additionally, there have been works on the effects of neutrino flavor conversions in the context of core-collapse supernovae~\cite{Ehring:2023lcd,Ehring:2023abs,Wang:2025nii}, and substantial discussion regarding the outcomes of fast flavor conversions using either radiation-only simulations~\cite{Xiong:2023vcm, Fiorillo:2024qbl, Fiorillo:2024qbl,Xiong:2024pue,Kost:2024esc,Liu:2024nku,Kneller:2024buy,Lacroix:2024pbb,Nagakura:2025brr}, or analyzing static snapshots from classical hydrodynamical simulations~\cite{Richers:2022dqa,Grohs:2022fyq,Froustey:2023skf,Grohs:2023pgq,Grohs:2025ajr}.

In this {\em Letter}, we consider, for the first time, the impact of rapid, lepton number-preserving neutrino-flavor equilibration that could result from quantum many-body interactions~\cite{Martin:2023gbo}, beyond standard model physics~\cite{Ehring:2024mjx}, or flavor instabilities \cite{Richers:2022zug} in neutron star mergers. We show that neutrino flavor transformation can impact the structure and composition of the remnant, possibly leaving detectable changes in the post-merger GW emission. Flavor transformation has a profound impact on the ejecta composition and associated nucleosynthesis yields, with relative isobaric abundances of some nuclear species changing by up to one order of magnitude.

\label{sec:setup}
\textit{Setup.---}We model neutrino-flavor transformation using a Bhatnagar-Gross-Krook (BGK) collisional operator~\cite{Bhatnagar:1954zz, Nagakura:2023jfi},
%\begin{equation}
$
1/\tau_a\left(f^a-f\right),
$
%\end{equation}
which models flavor conversion as a relaxation towards an asymptotic equilibrium state. In the previous equation, $\tau_a$ is the relaxation time,
% which should be consistent with the neutrino flavor conversion rate. 
$f$ is the neutrino distribution function, and $f^a$ denotes $f$ at its asymptotic mixing equilibrium state. Motivated by~\cite{Martin:2023gbo}, we determine the mixing equilibrium states assuming detailed balance for neutrino many-body interactions, that is the neutrino number densities are chosen such that the reactions $\nu_e \bar{\nu}_e \leftrightarrows \nu_\mu \bar{\nu}_\mu \leftrightarrows \nu_\tau \bar{\nu}_\tau$ are in equilibrium. Since only 4 neutrino species are considered, i.e. $\nu_e, \nu_x, \bar{\nu}_e, \bar{\nu}_x$, where $x$ stands for $\mu$ and $\tau$ (anti-)neutrinos combined, the assumption yields \reply{$4n^a_{e} n^a_{\overline{e}}=n^a_{x} n^a_{\overline{x}}$, where $n$ is the number densities of neutrinos.}

We perform BNS merger simulations with \texttt{THC\_M1}~\cite{Radice:2012cu, Radice:2013hxh, Radice:2013xpa, Radice:2021jtw}. Our code evolves the (lab frame) neutrino number density $n$, energy $E$ and flux $F^\alpha$ using conservation equations in the 3+1 form (see in~\cite{Radice:2021jtw}), where on the right hand side we add the BGK flavor conversion operators using the corresponding relaxation times and calculated equilibrium values\footnote{Note that the mixing equilibrium states of $E$ and $F^\alpha$ are determined by both the balance for neutrino number densities and energy conservations in the lab frame.}. \reply{For this work, we adopt the closure prescription and neutrino interaction rates described in the Supplemental Material. We do not consider neutrino-lepton inelastic scattering effects\footnote{We do, however, include pair processes with a simplified treatment based on the use of Kirchoff's laws (i.e., in the local thermodynamic equilibrium approximation), see details in~\cite{Radice:2021jtw}.}~\cite{Chiesa:2024lnu} in this work.} The neutrino mixing operators are implemented in an operator-split fashion. \reply{Note that, in our model, the neutrino flavor decoherence effects cannot be fully captured, since we do not evolve the off-diagonal components of the neutrino density matrix. Nevertheless, this model has been shown to quantitatively capture the relaxation to equilibrium of the fast flavor instabilities in local simulations~\cite{Nagakura:2023jfi}, and is adequate for the scenario we are considering here, in which detailed balance is quickly realized in flavor space.}

We construct initial data assuming non-rotating equal-mass neutron stars with mass $1.35$$M_{\odot}$ and an initial separation of $45$ km, corresponding to ${\sim}3{-}4$ orbits prior to merger,
using the Lorene pseudo-spectral code~\cite{Gourgoulhon:2000nn}. Neutron star matter is described using the DD2 equation of state~\cite{Typel:2009sy, Hempel:2009mc}. We employ 7 levels of adaptive mesh refinement~\cite{Berger:1984zza, Berger:1989a} using the Carpet  driver~\cite{Schnetter:2003rb, Reisswig:2012nc} of the Einstein Toolkit~\cite{Loffler:2011ay,roland_haas_2024_14193969}. The inner-most refinement level covers each star prior to merger, as well as the central part of the merger remnant. We present three sets of simulations, i) no neutrino mixing (`\texttt{No-Mixing}'), ii) neutrino mixing with density threshold $10^{11}$ g/cm$^3$ (`$\mathtt{\rho-11}$') and iii) neutrino mixing with density threshold $10^{13}$ g/cm$^3$ (`$\mathtt{\rho-13}$'). All simulations were conducted with both low resolution (LR) and standard resolution (SR), where the finest grid have spacing of $h=0.167\ M_{\odot} \simeq 246\ {\rm m}$ and $0.125\ M_{\odot} \simeq 184\ {\rm m}$, respectively. We discuss the SR results in the text. The LR results agree qualitatively, see Supplemental Material for more information. Similar to~\cite{Ehring:2023lcd, Ehring:2023abs}, in the $\mathtt{\rho-11}$ ($\mathtt{\rho-13}$) simulations, we use uniform relaxation times $\tau_a=0.4$ ns\footnote{A sub-nano second relaxation time is chosen to show that our scheme is capable of considering neutrino oscillations at fast flavor conversion timescale.} for regions where the rest mass density $\rho$ is below $10^{11}$$\mathrm{g/cm}^3$ ($10^{13}$$\mathrm{g/cm}^3$), and infinite relaxation times when $\rho$ is above the threshold. The $\mathtt{\rho-13}$ model is chosen because \reply{the approximate decoupling densities of heavy-lepton neutrinos are $10^{13}$$\mathrm{g/cm}^3$}. It is also roughly the density of the surface of the remnant~\cite{Hanauske:2016gia}, inside which the interaction with matter should dominate over the flavor conversion effects. The $\mathtt{\rho-11}$ model considers a case in which flavor transformations become important only in the optically thin region, \reply{as electron-type neutrinos also decouple} below density $10^{11}$$\mathrm{g/cm}^3$. \reply{Note that our simulations are subject to uncertainties due to finite-resolution effects~\cite{Zappa:2022rpd} and the use of moment-based scheme for neutrino transport~\cite{Foucart:2024npn}. However, the effects we report are larger than the estimated numerical uncertainties, which will be shown in the following text.}% We will report on a more exhaustive set of simulations in a forthcoming work.

\begin{figure*}
\includegraphics[width=2.1\columnwidth]{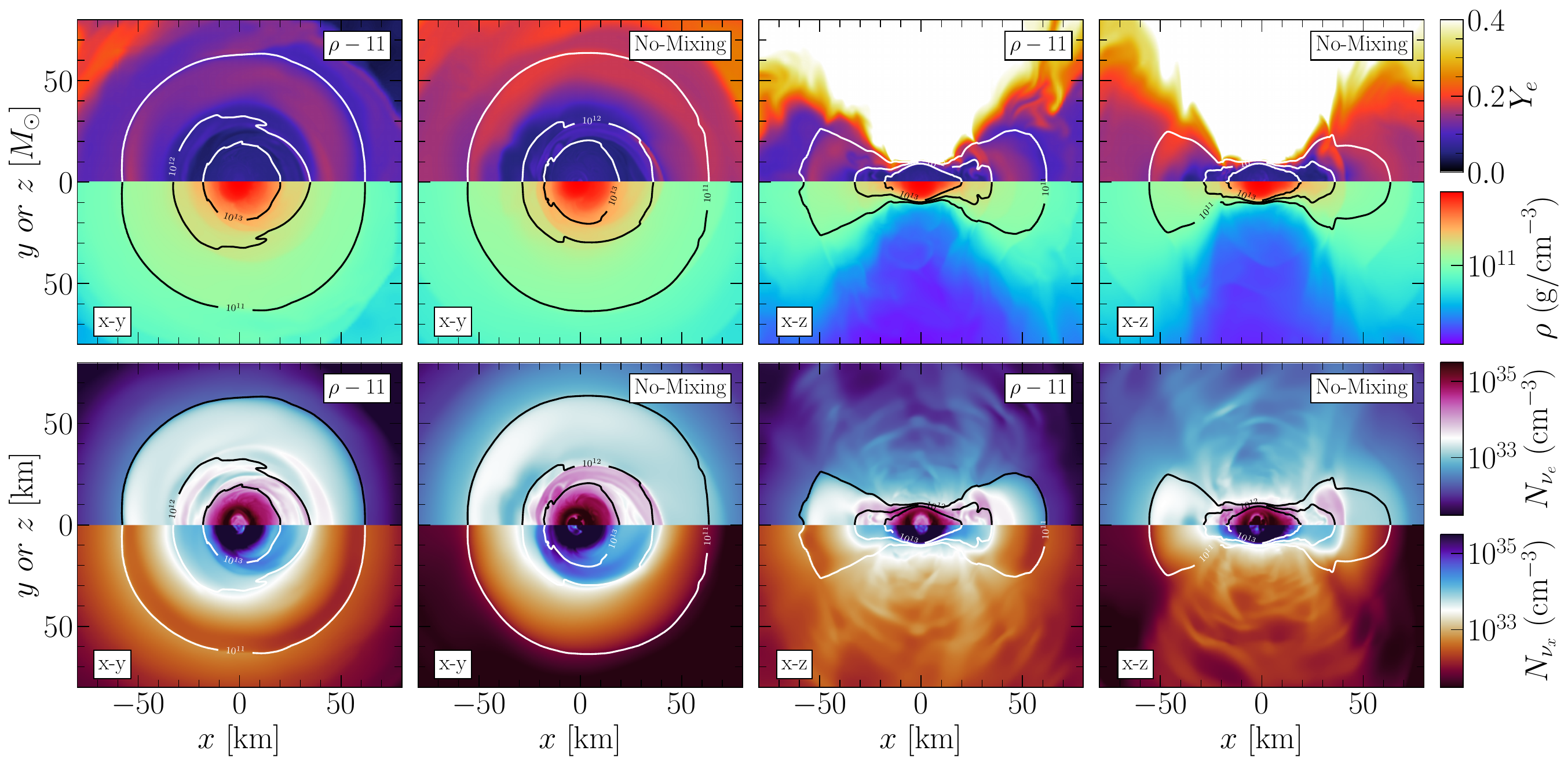}
\caption{Simulation snapshots in the equatorial: x-y (meridional: x-z) planes for the $\mathtt{\rho-11}$ and the \texttt{No-Mixing} models at $\sim$$15$ ms after merger on the left (right) four panels, respectively. The top-row panels show electron fractions (upper half) and rest mass densities (lower half). The second-row panels display electron neutrino number densities (upper half) and heavy lepton neutrino number densities (lower half). The contour lines denote where the density is at $10^{11}$, $10^{12}$ and $10^{13}$$\mathrm{g/cm}^3$. We find that after merger, the disk in the $\mathtt{\rho-11}$ mixing model is more neutron rich than that of the \texttt{No-Mixing} model. Due to flavor conversion effects transforming electron (anti-)neutrinos to heavy lepton neutrinos, we also find that electron neutrinos are more abundant in the \texttt{No-Mixing} than the $\mathtt{\rho-11}$, and vise versa for heavy lepton neutrinos.}
\label{fig:2d}
\end{figure*}

\begin{figure}[htbp]
\includegraphics[width=0.98\columnwidth]{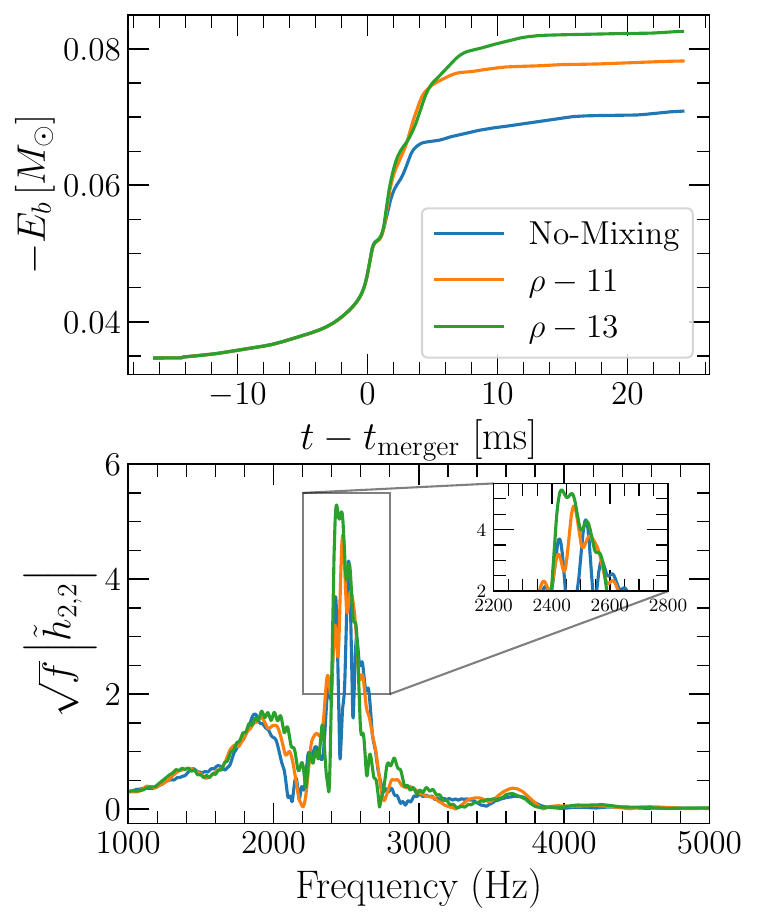}
\caption{Upper panel: evolution of the gravitational binding energy in the three models. The $\mathtt{\rho-13}$ remnant is the most bound, followed by the $\mathtt{\rho-11}$ simulation, and then by the \texttt{No-Mixing} simulation. Notably, the binding energy for the $\mathrm{\rho-13}$ remnant is $\sim$$20\%$ greater than the \texttt{No-Mixing} remnant. \reply{Lower panel: GW spectra in the three models. The post-merger peak frequencies shift from high to lower in the \texttt{No-Mixing}, the $\mathtt{\rho-11}$, and the $\mathtt{\rho-13}$ simulations. The difference between the peak frequency is of ${\sim}100$~Hz. However, while the trend in binding energy is robust with respect to grid resolution, the frequency shifts are not, so we cannot exclude that they are a numerical artifact.}}
\label{fig:energy}
\end{figure}

\begin{figure}
\includegraphics[width=0.98\columnwidth]{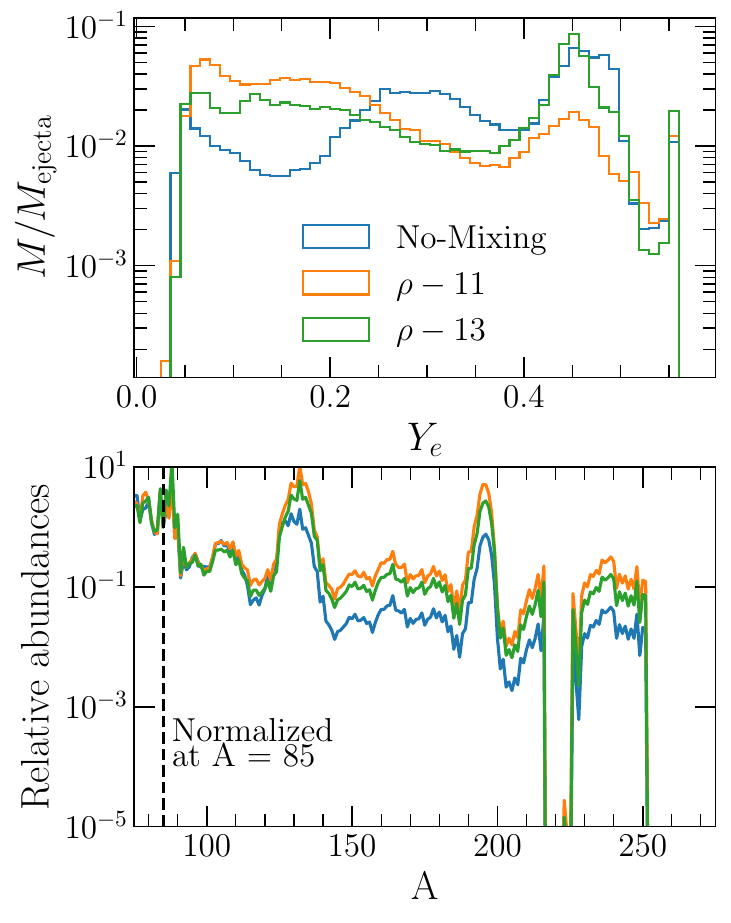}
    \caption{Upper panel: histograms of the electron fraction distributions of the ejecta. Due to neutrino mixing, the ejecta are generally more neutron rich in the $\mathtt{\rho-11}$ simulation and the $\mathtt{\rho-13}$ simulation than that in the \texttt{No-Mixing} simulation. The $\mathtt{\rho-11}$ simulation has the most low ($<0.2$) $Y_e$ material in the ejecta, followed by $\mathtt{\rho-13}$ simulation, and then by the \texttt{No-Mixing} simulation. Lower panel: the relative abundances of nuclei of mass number $A$ formed in the ejecta. We normalize the yields at $A=85$. Note that the yields for lanthanides and heavy elements are one order of magnitude higher in the $\mathtt{\rho-11}$ simulation than the \texttt{No-Mixing} simulation. Because the yields are determined by the r-process, which will be enhanced/suppressed by more/less neutron-rich ejecta, the ranking between the three models is consistent as the electron fraction histograms.} 
    \label{fig:ejecta}
\end{figure}

\begin{figure}
\includegraphics[width=0.98\columnwidth]{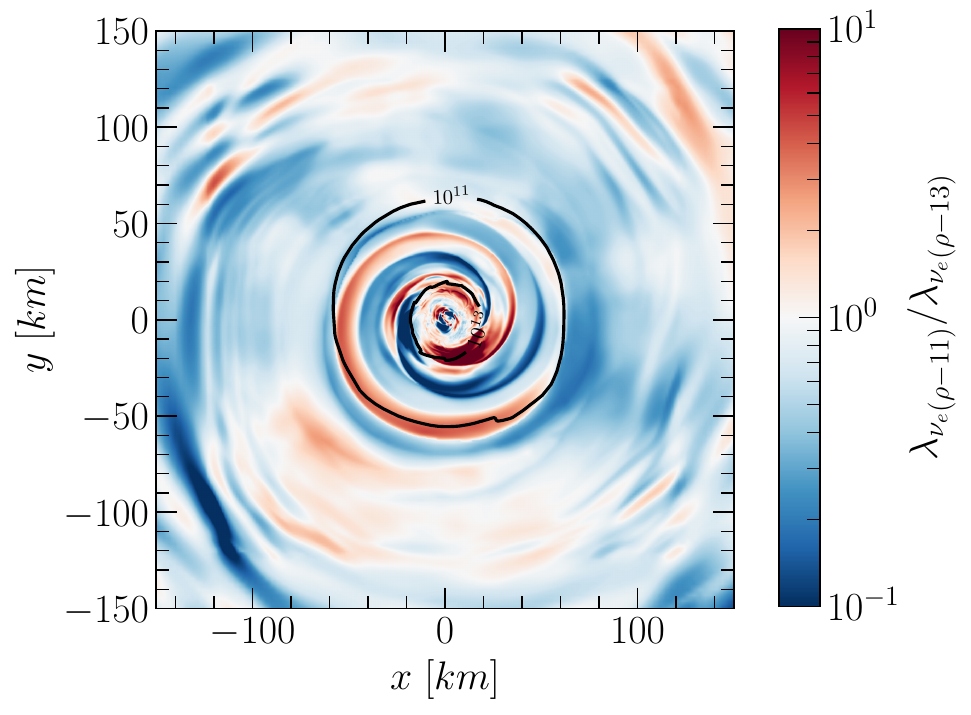}
    \caption{Ratio of the electron neutrino absorption rates for the $\mathtt{\rho-11}$ simulation over the $\mathtt{\rho-13}$ simulation on the x-y plane, taken at $\sim$$15$ ms after merger. The contours are drawn for rest mass density of $10^{11}$ and $10^{13}$$\mathrm{g/cm}^3$. We see greater electron neutrino absorption rates in the $\mathtt{\rho-13}$ simulation than those of the $\mathtt{\rho-11}$ simulation for the outer disk region where density is below $10^{11}$$\mathrm{g/cm}^3$.} 
    \label{fig:rate}
\end{figure}

\label{sec:results}
\textit{Results.---}
Our simulations start from the late inspiral phase of the BNS, the stars merge $\sim$$15$~ms from the beginning of the simulation. Since DD2 is a stiff equation of state the remnants are stable against gravitational collapse. We evolve the system for an additional $\sim$$30$ ms post-merger to be able to capture the bulk of the dynamical mass ejection. In Fig.~\ref{fig:2d}, we show the distribution of electron fractions, rest mass densities, electron neutrino and heavy lepton neutrino number densities on the x-y and x-z planes for the $\mathtt{\rho-11}$ and the \texttt{No-Mixing} simulations. These snapshots, taken at $\sim$$15$ ms post-merger, capture the general dynamics of the remnant disk. In the equatorial plane and lower angle polar regions, we observe that the electron fraction of the accretion disk surrounding the merger remnant is substantially lower in the $\mathtt{\rho-11}$ case than that in the \texttt{No-Mixing} simulation. \reply{In particular, $Y_e$ inside the $10^{11}\mathrm{g/cm}^3$ density contours also seem to be lower in the $\mathtt{\rho-11}$ case, possibly due to oscillations in the neutrino-flavor conversion surface due to the the spiral density waves \cite{Nedora:2019jhl}. Moreover, we observe mixing between the neutrino driven wind and the disk at intermediate latitudes, which might not be easily appreciated through the static snapshots in Fig.~\ref{fig:2d}.} In addition, as the electron (anti-)neutrinos undergoing $\nu_e, \bar{\nu}_e \rightarrow \nu_x, \bar{\nu}_x$ flavor conversions, we find electron neutrinos to be less abundant in the outer disk (i.e., $\rho<10^{11}$$\mathrm{g/cm}^3$) in the $\mathtt{\rho-11}$ simulation than the  \texttt{No-Mixing} simulation. Heavy lepton neutrinos exhibit the opposite trend.
% Despite these differences, the overall rest mass density distribution remains nearly identical between the two models, indicating that the large-scale dynamics are not significantly affected.

We compute the remnant binding energies in \reply{the upper panel of} Fig.~\ref{fig:energy}, by integrating the GW energy losses. The $\mathtt{\rho-13}$ simulation has the highest GW luminosity and the most bound remnant, followed by the $\mathtt{\rho-11}$ simulation, and then the least in the \texttt{No-Mixing} simulation. In particular, the binding energies for the $\mathrm{\rho-13}$ remnants are $\sim$$20\%$ greater than the \texttt{No-Mixing} remnants. Consistently, we find that these remnant have higher maximum density and smaller central value of the lapse function, as shown in the Supplementary Material. This could be the result of the enhanced thermal and electron (anti-)neutrino losses in the outer layers of these remnant favored by neutrino flavor transformation. We remark that a similar effect was recently reported by \cite{Ehring:2024mjx} in the context of core-collapse supernovae. This result suggests that gravitational wave observations could, at least in principle, constrain the physics of neutrino flavor transformation. We have also analyzed the post-merger GW spectra \reply{in the lower panel of Fig.~\ref{fig:energy}} and found evidence of a shift in the peak frequency of the post-merger signal as a result of mixing, \reply{with the discrepancy being up to 100 Hz}. However, \reply{we note here that such frequency shifts might be artificially exaggerated because of the double-peak feature seen in the \texttt{No-Mixing} spectrum. Moreover,} we could not resolve this trend in the lower-resolution simulations, so we cannot conclusively rule out that it is a numerical artifact.

To quantify the ejecta properties, we measure the flux of unbound material satisfying the geodesic criterion at a sphere located at $\sim$$200M_\odot$ ($295$ km) away from the center of the simulation domain. Fig.~\ref{fig:ejecta} compares the electron fraction profiles and nucleosynthesis yields \reply{at $\sim32$ years after merger~\footnote{We calculate the abundances by convolving ejecta properties with pre-computed nucleosynthesis trajectories computed with the SkyNet \cite{Lippuner:2015gwa, Lippuner:2017tyn} code. This procedure is discussed in more details and validated in \cite{Radice:2018pdn}.}} for the three models. Due to the flavor conversions from electron neutrinos to heavy lepton neutrinos, the induced deficits of electron neutrinos lead to a decrease in the neutrino reabsorption $\nu_e+n \rightarrow p+e^{-}$. Therefore, the mixing simulations have more neutron-rich ($Y_e<0.2$) materials in the ejecta, with the $\mathtt{\rho-11}$ simulation having the most, followed by $\mathtt{\rho-13}$ simulation, and the \texttt{No-Mixing} simulation the least. The $\mathtt{\rho-11}$ simulation generates up to 5 times more neutron-rich ejecta than the \texttt{No-Mixing} simulation, while $\mathtt{\rho-13}$ simulation produces up to twice as much. Both differences significantly exceed the reported $10\%-30\%$ numerical errors estimated from comparisons between M1 and Monte Carlo neutrino transport schemes~\cite{Foucart:2020qjb,Foucart:2024npn}. The lower panel of Fig.~\ref{fig:ejecta} presents the corresponding nucleosynthesis yields, calculated using the convolution method as in our previous work~\cite{Radice:2018pdn,Espino:2023mda}. Due to the strong dependence of r-process nucleosynthesis on the ejecta $Y_e$, we observe an order-of-magnitude difference in lanthanide and heavy element production between the $\mathtt{\rho-11}$ and the \texttt{No-Mixing} simulations, and a factor of two between the $\mathtt{\rho-13}$ and the \texttt{No-Mixing} simulations. Notably, the differences caused by flavor conversion effects range from approximately $200\%$ to $1000\%$. % This highlights the prominent impact of neutrino oscillation effects.

Furthermore, we find that changing the neutrino mixing conditions give rise to quantitatively different ejecta properties. In the $\mathtt{\rho-13}$ simulation, the mixing is turned on in the inner disk where density is between $10^{11}$$\mathrm{g/cm}^3$ and $10^{13}$$\mathrm{g/cm}^3$. In this region, electron (anti-)neutrinos remain in thermal equilibrium with the medium while undergoing flavor conversion to heavy lepton neutrinos. The latter can freely propagate outward. Once these heavy lepton neutrinos reach the outer disk where the density is below $10^{11}$$\mathrm{g/cm}^3$, a sufficient amount of them convert back into electron neutrinos, leading to an excess of electron neutrinos compared to the $\mathtt{\rho-11}$ simulation. Therefore, we see in Fig.~\ref{fig:ejecta}, the $\mathtt{\rho-13}$ simulation yields up to 3 times less neutron-rich ejecta, and correspondingly fewer heavy elements than the $\mathtt{\rho-11}$ simulation. 

Such effects are further illustrated in Fig.~\ref{fig:rate}, which shows the ratio of electron neutrino absorption rates, $\lambda_{\nu_{e}}$ between the $\mathtt{\rho-11}$ and the $\mathtt{\rho-13}$ simulations. In neutron-rich ejecta, the electron neutrino absorption rate is approximately proportional to the electron fraction equilibration rate~\cite{Qian:1996xt,Martin:2017dhc,Cusinato:2021zin}
\begin{equation}
    \frac{\mathrm{d} Y_e}{{\mathrm{~d} t}}=\lambda_{\nu_{\mathrm{e}}}\left(1-Y_{\mathrm{e}}\right)-\lambda_{\bar{\nu}_{\mathrm{e}}} Y_{\mathrm{e}}\approx\lambda_{\nu_{\mathrm{e}}}.
\end{equation}
In the outer disk where density is below $10^{11}$$\mathrm{g/cm}^3$, the electron neutrino absorption rate is significantly higher in $\mathtt{\rho-13}$ simulation than those of the $\mathtt{\rho-11}$ simulation, leading to a more rapid increase in $Y_e$ for the corresponding ejecta. This explains the less neutron-rich ejecta and heavy element produced in the $\mathtt{\rho-13}$ simulation. 

\label{sec:conclusion}
\textit{Conclusion.---}We report on the first numerical relativity simulations including the dynamical effects of lepton number-preserving neutrino flavor equilibration that could be driven by quantum many-body effects, beyond standard-model physics, or flavor instabilities. Our study demonstrates that such transformations can impact the structure and composition of the merger remnant and of its accretion disk. At high densities, the conversion of electron-type neutrinos and anti-neutrinos into heavy-lepton neutrinos accelerates the contraction of the outer layers of the remnant. This is manifested by a significant increase, of up to ${\sim}20$\%, in the overall postmerger GW luminosity, which could be probed with future GW experiments, such as Cosmic Explorer \cite{Reitze:2019iox} and the Einstein Telescope \cite{Punturo:2010zz}. At lower densities, the depletion of electron-type neutrinos due to flavor conversion results in more neutron rich ejecta and boosts the production of heavy r-process elements with $A \gtrsim 120$ by up to one order of magnitude. \reply{We note that such differences are results of our neutrino flavor transformation treatments and choice of mixing prescriptions. In different flavor transformation scenarios, e.g., when considering the effects of the fast-flavor instability, the quantitative impact might change.}

Interestingly, not only whether or not neutrino flavor transformation occurs, but also \emph{where} it occurs has a strong impact on the nucleosynthesis. In particular, we examine two density-dependent neutrino mixing models where neutrino flavor conversions are considered everywhere with density below either $10^{11}$ or $10^{13}$$\mathrm{g/cm}^3$. While both cases have significantly more neutron rich ejecta and higher abundances of heavy r-process elements than the $\texttt{No-Mixing}$ runs, there are important differences between them. In particular, if neutrinos undergo flavor transformation within the disk, e.g., at densities between $10^{11}$ and $10^{13}$$\mathrm{g/cm}^3$, this boosts the neutrino losses from the disk, because the disk is optically thin to heavy lepton-neutrinos. Some of the escaping neutrinos are then converted back to electron-type neutrinos and absorbed into the ejecta, decreasing the neutron richness and $\langle A \rangle$ of the ejecta.

Future work will extend this work by exploring mixing prescriptions motivated by the fast-flavor instability~\cite{Abbar:2024ynh, Wang:2025nii} and considering a broader range of equations of states and merger outcomes. \reply{We also aim to include muonic interactions\cite{Pajkos:2024iml,Ng:2024zve}, along with kernels for neutrino-lepton inelastic scatterings and pair processes~\cite{Cheong:2024cnb,Chiesa:2024lnu}, which are missing in this work and other published BNS simulations.}

\vspace{1em}
\begin{acknowledgements}
It is our pleasure to acknowledge Joseph Carlson, Gail McLaughlin, Hiroki Nagakura, and Albino Perego for discussions during various stages of development of this work.
DR, SR, and MB thank the Institute for Nuclear Theory at the University of
Washington for the kind hospitality during the INT-23-2 program, during which this project was conceived. This research was supported in part by the INT's U.S. Department of Energy grant No. DE-FG02- 00ER41132.
YQ and MB were supported by the U.S. Department of Energy, Office of Science,
Division of Nuclear Physics under Award Number(s) DE-SC0021177 and DE-SC0024388
DR acknowledges support from the Sloan Foundation, from the National Science
Foundation under Grants No. PHY-2020275, AST-2108467, PHY-2116686, and
PHY-2407681. SR acknowledges support from the National Science Foundation under Grant No. PHY-2412683.
Simulations were performed on TACC's Frontera (NSF LRAC allocation
PHY23001), on NERSC's Perlmutter, and on the Pennsylvania State University’s
Institute for Computational and Data Sciences’ Roar supercomputer. This research
used resources of the National Energy Research Scientific Computing Center, a
DOE Office of Science User Facility supported by the Office of Science of the
U.S.~Department of Energy under Contract No.~DE-AC02-05CH11231.
\end{acknowledgements}

%%%%%%%%%%%%%%%%%%%%%%%%%%%%%%%%%%%%%%%%%%%%%%%%%%%%%%%%%%%%%%%%%%%%%%
\bibliography{draft.bib}
%%%%%%%%%%%%%%%%%%%%%%%%%%%%%%%%%%%%%%%%%%%%%%%%%%%%%%%%%%%%%%%%%%%%%
%%%%%%%%%%%%%%%%%%%%%%%%%%%%%%%%%%%%%%%%%%%%%%%%%%%%%%%%%%%%%%%%%%%%%
\clearpage
\appendix
\section{SUPPLEMENTAL MATERIAL}
\label{App:Supplemental analysis}

% \subsection{Neutrino Luminosities}

\subsubsection{The \texttt{THC\_M1} Scheme}
\label{sec:closure}

In \texttt{THC\_M1}, we decompose the (neutrino) radiation stress energy tensors as
\begin{equation}
T^{\alpha \beta}_{\rm{rad}}=E n^\alpha n^\beta+F^\alpha n^\beta+n^\alpha F^\beta+P^{\alpha \beta},
\end{equation}
where $n^\alpha$ is the future-oriented unit vector normal to the constant time hypersurfaces. The decomposition also gives $P^{\alpha \beta}$, the radiation pressure tensor, as the second order moment.
In order to close the $E$ and $F^\alpha$ equations, we need to relate $P^{\alpha \beta}$ with $E$ and $F^\alpha$. A standard prescription is to use an analytic closure to express the neutrino radiation pressure in different optical depth regimes~\cite{Shibata:2011kx}
\begin{equation}
P_{\alpha \beta}=\frac{3 \chi-1}{2} P_{\alpha \beta}^{\text {thin }}+\frac{3(1-\chi)}{2} P_{\alpha \beta}^{\text {thick }},
\end{equation}
where $\chi \in\left[\frac{1}{3}, 1\right]$ is the Eddington factor. Using Minerbo closure \cite{MINERBO}, we express $\chi$ as
\begin{equation}
\chi(\xi)=\frac{1}{3}+\xi^2\left(\frac{6-2 \xi+6 \xi^2}{15}\right),\quad\xi^2=\frac{\tilde{F}_\alpha \tilde{F}^\alpha}{\tilde{E}^2}
\end{equation}
where the $\tilde{\cdot}$ quantities are the radiation fields in the fluid co-moving frame. In this work, we use the following ansatz for $P^{\text {thick }}$
\begin{equation}
\tilde{P}^{\text {thick }}_{\alpha \beta}=\frac{1}{3}\tilde{E}\left(g_{\alpha \beta}+u_\alpha u_\beta\right)
\end{equation}
where $u^\alpha$ is the fluid four-velocity, $g_{\alpha \beta}$ is the spacetime metric. In addition, we use
\begin{equation}
P^{\text {thin }}_{\alpha \beta}=\frac{F_\alpha F_\beta}{E}
\end{equation}
in the optically thin limit~\cite{Shibata:2011kx, Murchikova:2017zsy}.
\begin{replies}

The weak reactions considered in this work are shown in Table.~\ref{tab:weak}, which are the same as those in~\cite{Radice:2018pdn}. The formulas from the references listed in the table are used to compute the neutrino emissivity, absorption and scattering rates.

\begin{table}[htbp]
\centering
\caption{Weak reactions considered in this work and references for their implementation. In the reaction expressions, N denotes a nucleon, and A denotes a nucleus.}
\label{tab:weak}
\begin{tabular}{ll}
\hline \hline
Reaction & Reference \\
\hline
$\nu_e + n \leftrightarrow p + e^{-}$ & \cite{Bruenn:1985en} \\
$\bar{\nu}_e + p \leftrightarrow n + e^{+}$ & \cite{Bruenn:1985en} \\
$e^{+} + e^{-} \rightarrow \nu + \bar{\nu}$ & \cite{Ruffert:1995fs} \\
$\gamma + \gamma \rightarrow \nu + \bar{\nu}$ & \cite{Ruffert:1995fs} \\
$N + N \rightarrow \nu + \bar{\nu} + N + N$ & \cite{Burrows:2004vq} \\
$\nu + N \rightarrow \nu + N$ & \cite{Ruffert:1995fs} \\
$\nu + A \rightarrow \nu + A$ & \cite{Shapiro:1983du} \\
\hline \hline
\end{tabular}
\end{table}

\begin{figure}
\includegraphics[width=0.98\columnwidth]{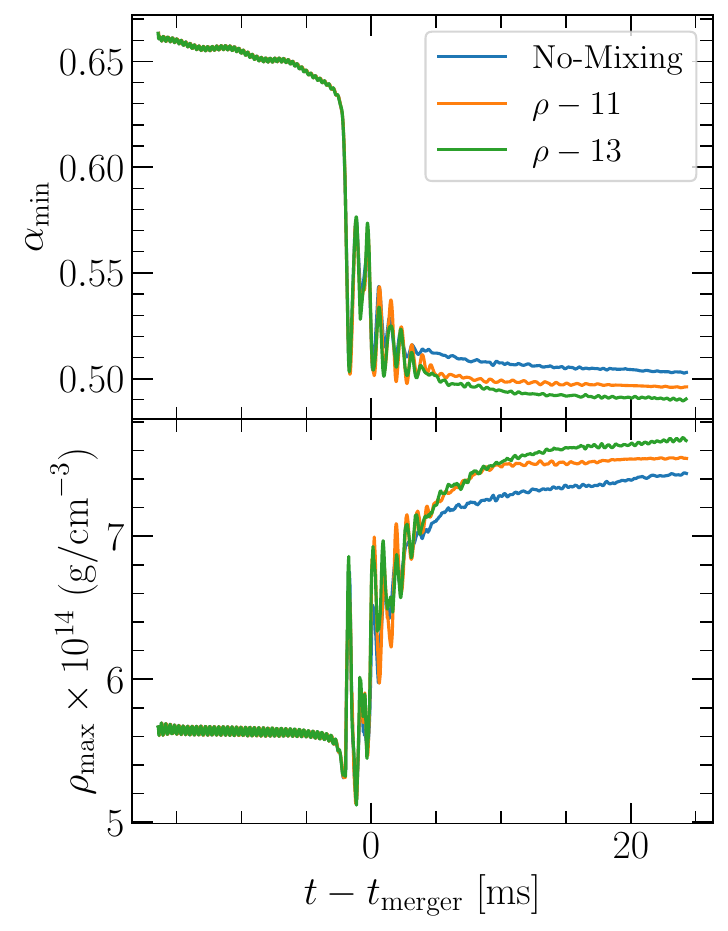}
\caption{The evolutions of the minimum lapse (upper), and maximum density (lower) in the three models. The remnant NSs become stable at $\sim$10 ms after merger, we see the highest compactness in the $\mathtt{\rho-13}$ simulation, then the $\mathtt{\rho-11}$ simulation, and the least compact remnant in the \texttt{No-Mixing} simulation. The relative differences between different models are within $5\%$.}
\label{fig:2mass}
\end{figure}
\subsubsection{Neutrino Flavor Conversion}
\begin{figure*}
\includegraphics[width=2.0\columnwidth]{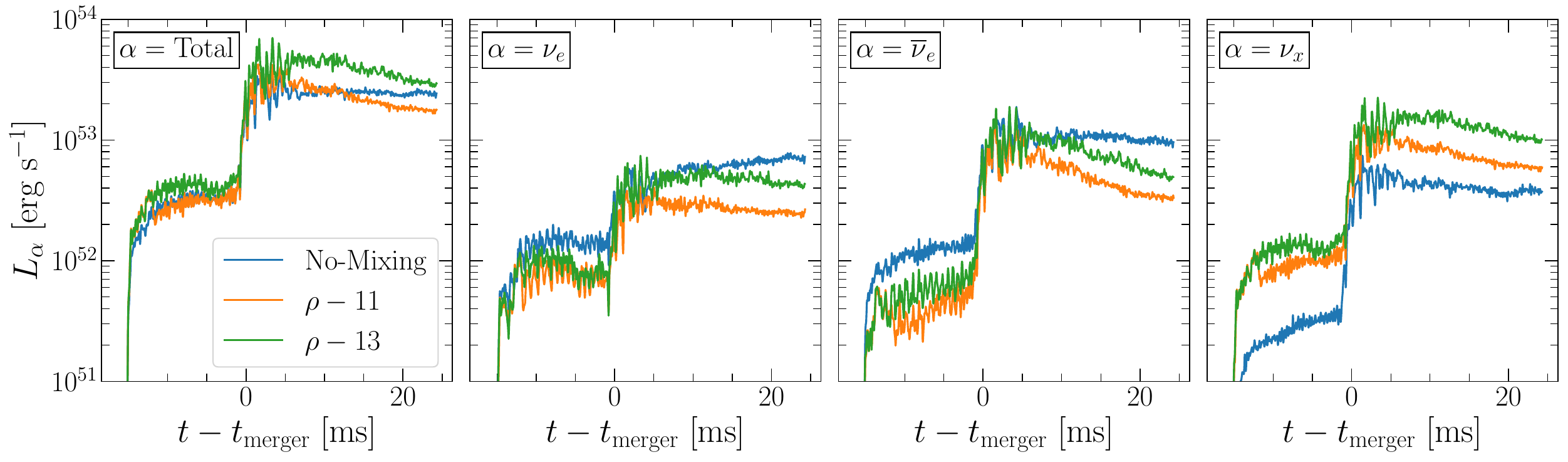}
        \caption{Neutrino luminosities evolutions for total, $\nu_e$, $\overline{\nu}_e$, and $\nu_x$, respectively, in panels from left to right. Driven by the flavor conversions $\nu_e, \bar{\nu}_e \rightarrow \nu_x, \bar{\nu}_x$, we see that electron (anti-)neutrinos are more luminous in the \texttt{No-Mixing} simulation than the $\mathtt{\rho-11}$ and the $\mathtt{\rho-13}$ simulations, vise versa for heavy lepton neutrinos. Between two mixing models, the $\mathtt{\rho-13}$ simulation has higher luminosities for all neutrino species compared to the $\mathtt{\rho-11}$ simulation, which are due to the additionally enabled flavor conversion effects between $10^{11}$$\mathrm{g/cm}^3$ and $10^{13}$$\mathrm{g/cm}^3$ inner disk region.}
        \label{fig:lumi}
\end{figure*}

We introduce neutrino mixing in the lab frame \emph{after} the classical source term updates and neutrino fields feed back to the fluid. The mixing effect is applied using the BGK operator on the energy and flux components
\begin{equation}
\begin{aligned}
&n^\text{new} = \lambda n^\text{old} + (1 - \lambda) n^\text{mix},\\
&E^\text{new} = \lambda E^\text{old} + (1 - \lambda) E^\text{mix},\\
&F_i^\text{new} = \lambda F_i^\text{old} + (1 - \lambda) F_i^\text{mix},
\end{aligned}
\label{eq:lambda}
\end{equation}
where $n$ is the number density of neutrinos in the lab frame, while \(\lambda = e^{-\Delta t / \tau}\). In the limit \(\Delta t \to \infty\), the new lab frame neutrino number density, energy density and flux components approach their mixing equilibrium values.

To determine the mixing equilibrium values of energies and fluxes for different neutrino species, we first compute the mixing equilibrium number densities. In our current implementation, only four neutrino species are evolved, with heavy lepton species combined. The neutral current neutrino-neutrino interaction terms that drive flavor instabilities require local conservation of the following quantities:
\begin{equation}
\label{eq:conserved_quantities}
\begin{aligned}
N &= n_e + n_x + \bar{n}_e + \bar{n}_x, \\
N_e &= n_e - \bar{n}_e, \\
N_x &= n_x - \bar{n}_x.
\end{aligned}
\end{equation}
where the subscripts of $n$ denote the corresponding (anti-)neutrino species, e.g., $n_e$ denotes the number density of electron neutrinos. Expecting a flavor transformation equilibrium defined by $n_e\bar{n}_e=n_\mu \bar{n}_\mu=n_\tau\bar{n}_\tau$, the number densities approach
\begin{equation}
\begin{aligned}
n_e^\text{mix} \, \bar{n}_e^\text{mix} = \frac{1}{4} n_x^\text{mix} \, \bar{n}_x^\text{mix}.
\end{aligned}
\end{equation}
Letting \(x = n_e^\text{mix}\), we have:
\begin{equation}
\begin{aligned}
\bar{n}_e^\text{mix} &= x - N_e, \\
\bar{n}_x^\text{mix} &= n_x^\text{mix} - N_x,
\end{aligned}
\end{equation}
and from total number conservation:
\begin{equation}
\begin{aligned}
N &= x + \bar{n}_e^\text{mix} + n_x^\text{mix} + \bar{n}_x^\text{mix} \\
&= x + (x - N_e) + n_x^\text{mix} + (n_x^\text{mix} - N_x),
\end{aligned}
\end{equation}
which implies
\begin{equation}
n_x^\text{mix} = \frac{1}{2}(N + N_e + N_x) - x.
\end{equation}
Substituting into the equilibrium condition gives
\begin{equation}
\begin{aligned}
x\left(x-N_e\right)=&\frac{1}{4}\left[\frac{1}{2}\left(N+N_e+N_x\right)-x\right] \\
& \left\{\left[\frac{1}{2}\left(N+N_e+N_x\right)-x\right]-N_x\right\},
\end{aligned}
\end{equation}
Solving this equation, we find
\begin{equation}
n_e^\text{mix} = x = -\frac{N}{6} + \frac{N_e}{2} + \frac{\sqrt{4 N^2 + 12 N_e^2 - 3 N_x^2}}{6}.
\end{equation}
The other species can then also be determined from the conserved quantities in Eq.~\ref{eq:conserved_quantities}. Note that this prescription naturally preserves the positivity of the neutrino number densities and satisfies conservation of total lepton number, electron lepton number (ELN), and heavy lepton number (XLN). Once the equilibrium number densities are determined, we define a \(4 \times 4\) mixing matrix \(Y_{a,b}\), such that the diagonal components of \( Y \) are given by
\begin{equation}
Y_{a,a} = \min\left(1, \frac{n_a^\text{mix}}{n_a}\right),
\end{equation}
for \( a = 0,1,2,3 \) correspond to electron neutrinos, electron anti-neutrinos, heavy lepton neutrinos and heavy lepton anti-neutrinos, respectively. The off-diagonal components redistribute the remaining number densities to preserve total lepton numbers
\begin{equation}
\begin{aligned}
Y_{2,0} &= 1 - Y_{0,0}, & \text{(}\nu_e \rightarrow \nu_x\text{)} \\
Y_{3,1} &= 1 - Y_{1,1}, & \text{(}\bar{\nu}_e \rightarrow \bar{\nu}_x\text{)} \\
Y_{0,2} &= 1 - Y_{2,2}, & \text{(}\nu_x \rightarrow \nu_e\text{)} \\
Y_{1,3} &= 1 - Y_{3,3}, & \text{(}\bar{\nu}_x \rightarrow \bar{\nu}_e\text{)}
\end{aligned}
\end{equation}
All other components of \( Y \) are set to zero:
\begin{equation}
Y_{a,b} = 0, \quad \text{for all other } (a,b).
\end{equation}
The mixed number densities then satisfy
\begin{equation}
n_a^\text{mix} = Y_{a,b} n_b\,,
\end{equation}
\begin{figure*}
\includegraphics[width=2.0\columnwidth]{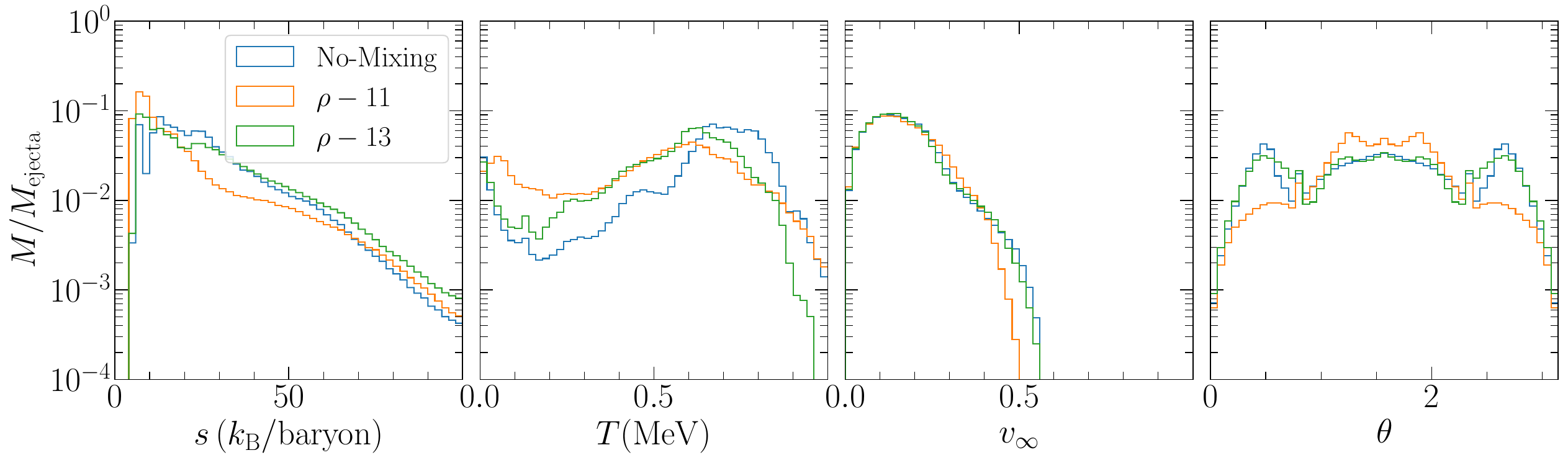}
\caption{Histograms of the dynamical ejecta entropy, temperature, velocity at infinity and polar angle. The ejecta are observed on a sphere located at $\sim$$200 M_\odot$ ($295$ km) away from the center of the simulation domain, based on geodesic criterion. The \texttt{No-Mixing} simulation, the $\mathtt{\rho-11}$  simulation and the $\mathtt{\rho-13}$ simulation ejecta are drown in blue, orange and green, respectively. All diagnostics look similar in three models.}
\label{fig:4hist}
\end{figure*}
and the structure of $Y_{ab}$ enforces the conservation laws in Eq.~\ref{eq:conserved_quantities}. Finally, we apply the same mixing transformation to energy and flux (separately for each direction)
\begin{equation}
E_a^\text{mix} = Y_{a,b} E_b, \qquad F_a^\text{mix} = Y_{a,b} F_b.
\end{equation}
and update the fields according to Eq.~\ref{eq:lambda}.
\end{replies}
\subsubsection{Additional Diagnostics}
\label{sec:additional}
In the top panel of Fig.~\ref{fig:2mass} we show the minimum lapse as a function of time for the \texttt{No-Mixing}, the $\mathtt{\rho-11}$, and the $\mathtt{\rho-13}$ simulations. As a consequence of our choice of a stiff equation of state, we see stable post-merger remnants for all simulations. Since remnant central density is inversely correlated with the minimum lapse, we find that the $\mathtt{\rho-13}$ simulation remnant is the most compact, followed by $\mathtt{\rho-11}$ simulation, with the \texttt{No-Mixing} case being the least compact. Their relative differences are more than $5\%$. This trend is further confirmed by the evolution of the maximum density, $\rho_\mathrm{max}$, as shown in bottom panel of Fig.~\ref{fig:2mass}. The ranking of compactness, the $\mathtt{\rho-13}$ simulation, followed by $\mathtt{\rho-11}$ simulation, and then the \texttt{No-Mixing} simulation, is consistent in both diagnostics. This may be due to the flavor swapping induced electron (anti-)neutrino losses in the outer layers of the remnant.

% \begin{figure}
% \includegraphics[width=0.98\columnwidth]{Fig/1gw_psd.pdf}
% \caption{The GW spectra in the three models. The post-merger peak frequencies shift from high to lower in the \texttt{No-Mixing}, the $\mathtt{\rho-11}$, and the $\mathtt{\rho-13}$ simulations. The difference between the peak frequency can be up to 100 Hz.}
% \label{fig:gw}
% \end{figure}

We keep track of the neutrino fluxes passing through a sphere of $\sim$$295$ km radius from the center of simulation domain and integrate them to obtain for--overall and individual flavors--neutrino luminosities. As shown in Fig.~\ref{fig:lumi}, in post-merger phase, the total neutrino luminosities are highest in the $\mathtt{\rho-13}$ simulation, with the other two simulations being qualitatively comparable. For individual species, we see $\mathtt{\rho-13}$ simulation has up to $400\%$ more heavy lepton neutrino luminosity than the \texttt{No-Mixing} simulation, and up to $100\%$ more than the $\mathtt{\rho-11}$ simulation. In contrast, the electron (anti-)neutrinos luminosities are down by up to $200\%$ in the $\mathtt{\rho-11}$ simulation compared to the \texttt{No-Mixing} simulation. The results suggest that flavor conversions of $\nu_e, \bar{\nu}_e \rightarrow \nu_x, \bar{\nu}_x$ are significant enough to be observed. Moreover, the fact that different mixing conditions lead to different neutrino luminosities also suggest that \emph{where} neutrino flavor conversions occur is crucial for relevant observations.

In addition to electron fraction, we also show in Fig.~\ref{fig:4hist} for histograms of other relevant ejecta properties for SR simulations, including entropy, temperature, velocity at infinity and polar angle in the \texttt{No-Mixing}, the $\mathtt{\rho-11}$, and the $\mathtt{\rho-13}$ models. The ejecta entropy distribution seems to be almost the same across the models, all peaked at $\sim$$10$ $k_{\mathrm{B}} / \text {baryon}$. Temperature and polar angle distributions have broad distributions. The ejecta velocities are lower than $0.5$ in all simulations. In general, despite minor fluctuations due to numerical uncertainties, the overall properties of the ejecta, other than $Y_e$, are similar across our simulations. 

\begin{figure}[b]
\includegraphics[width=0.98\columnwidth]{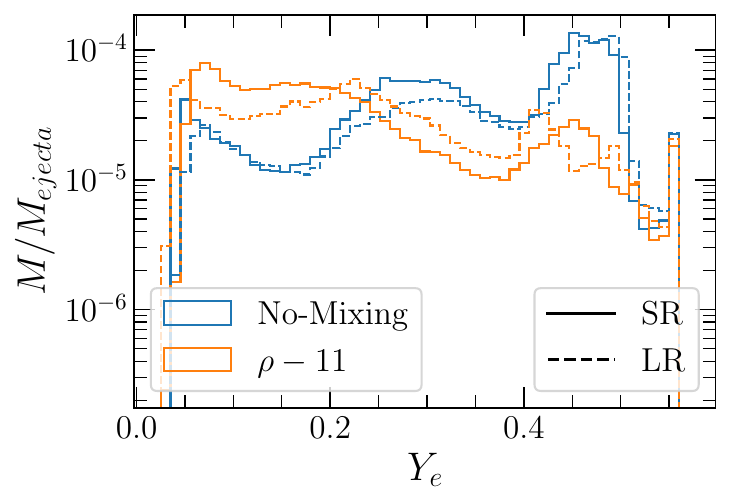}
\caption{The electron fraction distributions of the dynamical ejecta in the \texttt{No-Mixing} and the $\mathtt{\rho-11}$ simulations for both standard (solid) and low (dashed) resolutions. We see in both LR and SR that, the ejecta is generally neutron-richer in the $\mathtt{\rho-11}$ simulation than that in the \texttt{No-Mixing} simulation, showing good consistency against resolutions.}
\label{fig:SRvsLR}
\end{figure}

\subsubsection{Finite-Resolution Effects}

As a consistency test, we compare the electron fraction profiles of the dynamical ejecta in Fig.~\ref{fig:SRvsLR} for the \texttt{No-Mixing} and the $\mathtt{\rho-11}$ models in different resolutions. We see qualitatively the same distributions for the $Y_e$ in corresponding LR vs. SR comparisons. The relative difference in neutron-rich tails does not exceed $30\%$ for the $\mathtt{\rho-11}$ LR and SR simulations, while the \texttt{No-Mixing} LR and SR simulations show negligible differences as well. 

\end{document}